\documentclass[a4paper]{article}
\usepackage{graphicx}
\usepackage{url}  % Formatting web addresses
\usepackage{multicol}   %Columns
\usepackage[noae]{Sweave}
\usepackage{color} % text color for correcting manuscript
\usepackage{amsmath}
\usepackage{hyperref}
\usepackage{natbib}
\usepackage{amssymb}
 \usepackage{mathptmx}      % use Times fonts if available on your TeX system
\let\code=\texttt
\let\proglang=\textsf
\newcommand{\pkg}[1]{{\fontseries{b}\selectfont #1}}

\newcommand{\X}{\mathbf{X}}

\newcommand{\x}{\mathbf{x}}
\newcommand{\bb}{\boldsymbol{\beta}}
\newcommand{\bv}{\hat{\boldsymbol{\beta}}}

\begin{document}

\title{Identification of Outlying Observations with Quantile Regression for Censored Data%\thanks{Grants or other notes
}
%\subtitle{An add-on package for \proglang{R}}
%\titlerunning{Outlier Detection for Censored Data}        % if too long for running head

\author{Soo-Heang Eo\footnote{Department of Statistics, Korea University, Seoul, South Korea, 136-701,  hanansh@korea.ac.kr}       \and
        Seung-Mo Hong\footnote{Department of Pathology, Asan Medical Center, Seoul, South Korea, smhong28@gmail.com} \and
	HyungJun Cho\footnote{Corresponding author, Department of Statistics, Korea University, Seoul, South Korea, 136-701,  hanansh@korea.ac.kr} %etc.
}

%\authorrunning{Eo et al.} % if too long for running head

\maketitle

%\institute{Soo-Heang Eo \and
   %           Department of Statistics, Korea University, Seoul, South Korea, 136-701 \\
      %        \email{hanansh@korea.ac.kr}           %  \\
             %\emph{Present address:} of F. Author  %  if needed
          % \and
           %Seung-Mo Hong \and
              %Department of Pathology, Asan Medical Center, Seoul, South Korea, \\
              %\email{smhong@gmail.com}           %  \\
	%\and
	 %HyungJun Cho \and
             % Department of Statistics, Korea University, Seoul, South Korea, 136-701 \\
              %\email{hj4cho@korea.ac.kr}           %  \\
             %\emph{Present address:} of F. Author  %  if needed
%}


\begin{abstract}
Outlying observations, which significantly deviate from other measurements, may distort the conclusions of data analysis.
Therefore, identifying outliers is one of the important problems that should be solved to obtain reliable results.
While there are many statistical outlier detection algorithms and software programs for uncensored data, few are available for censored data.
In this article, we propose three outlier detection algorithms based on censored quantile regression, two of which are modified versions of existing algorithms for uncensored or censored data, while the third is a newly developed algorithm to overcome the demerits of previous approaches.
The performance of the three algorithms was investigated in simulation studies.
In addition, real data from SEER database, which contains a variety of data sets related to various cancers, is illustrated to show the usefulness of our methodology.
The algorithms are implemented into an \proglang{R} package \pkg{OutlierDC} which can be conveniently employed in the \proglang{R} environment and freely obtained from CRAN.

\end{abstract}

%%%%%%%%%%%%%%%%%%%%%%%%%%%%%%%%%%%%%%%%%
\section{Introduction}
\label{intro}
\qquad In various medical studies, outcomes of interest include the time to death or time to tumor recurrence.
When observing survival times, censoring is common because of partially known information.
Thus, survival data consist of survival time as an outcome, censoring status, and many covariates as risk factors.
The relationship between survival time and the covariates has been studied extensively.
In survival data analysis, customized statistical methodologies are employed because of non-normal distributions and  censoring.
The Surveillance, Epidemiology and End Results (SEER) Program, a premier source for cancer statistics in the United States, contains information on incidence, prevalence, and survival from specific geographic areas representing 28 percent of the US population \citep{hankey1999surveillance}.
Survival as an endpoint is one of the important outcomes of the SEER database; hence, it is often of interest to determine the relationship between survival and covariates.
When analysing SEER data, it is important to screen for outlying observations or outliers that significantly deviate from other measurements because they may distort the conclusions of the data analysis \citep{MR2112740}.
Therefore, the development of outlier detection methods is essential to obtain reliable results.

Outlier detection has been studied in various types of data including normal data \citep{hoaglin1986performance, chaloner1988bayesian, sim2005outlier}, multivariate normal data \citep{schwager1982detection}, censored data \citep{nardi1999new}, incomplete survey data \citep{beguin2004multivariate}, time series data \citep{atkinson2010forward}, gene expression data \citep{alshalalfa2012detecting}, proteomics data \citep{eo2012outlier, cho2008outlierd}, functional data \citep{sawant2012functional}, spatial data \citep{lu2003algorithms,sun2011functional} and circular data \citep{abuzaid2012boxplot}.
For more details, see \cite{MR2979735}, \cite{han2013outlier}, and \cite{aggarwal2013outlier}.  The algorithm of \cite{nardi1999new} was developed to identify outlying observations based on Cox linear regression for censored data. It can be more effective to utilize quantile regression because of it is robust to outliers. \cite{cho2008outlierd} and \cite{eo2012outlier} proposed to use quantile regression for outlier detection in proteomics data. The most algorithms focus on determining whether observations are outliers according to a threshold, which should be specified in advance.
The dichotomous algorithms, which depend solely on a pre-specified cut-off, may often be unsatisfactory. Thus, a function for providing scores and flexibly determining a threshold can be helpful.

In this paper, we present three outlier detection algorithms for censored data: the residual-based, boxplot, and scoring algorithms.
The residual-based and boxplot algorithms were developed by modifying existing algorithms \citep{nardi1999new, sas2008sas, eo2012outlier}, and the scoring algorithm was developed to provide the outlying magnitude of each point from the distribution of observations and enable the determination of a threshold by visualizing the scores.
The presented algorithms are based on quantile regression, which is robust to outliers.
The algorithms were investigated by a simulation study, and their characteristics were summarized at the conclusion of that study.
We implemented the three algorithms customized for censored survival data in an \proglang{R} package called \pkg{OutlierDC}, which can be conveniently employed in the \proglang{R} environment and freely obtained from the Comprehensive R Archive Network (CRAN) website (\url{http://cran.r-project.org/web/packages/OutlierDC/index.html}).
We demonstrate its use with real data from the SEER database (\url{http://seer.cancer.gov}), which contains a number of data sets related to various cancers.

The remainder of this paper is organized as follows.
In Sections \ref{theory} and \ref{package}, we describe three algorithms using censored quantile regression for identifying outlying observations in censored data and then implement them into an \proglang{R} package  \pkg{OutlierDC}.
In Section \ref{simul}, simulation studies are conducted to investigate the performance of the outlier detection algorithms.
In Section \ref{ex}, we illustrate the application of the algorithm using \pkg{OutlierDC} with a real example.
We present our conclusions in Section \ref{dis}.

%%%%%%%%%%%%%%%%%%%%%%%%%%%%%%%%%%%%%%%%%%%%%%%%%%%%%%%%%%%%%%%%%%%%%%%%%%%%%%%%%%%%%%%
\section{Outlier Detection for Censored Data}
\label{theory}
\qquad In this section, we describe three outlier detection algorithms based on censored quantile regression.
We here focus only on detecting too large observations, because too small observations can be generated by censoring.

We first define the notation used to explain the algorithms.
Let $T_{i}$ be an uncensored dependent variable of interest, such as survival time or some transformation of it, and let $C_{i}$ and $\X_{i}$ be a censoring variable and a $p$-dimensional covariate vector for the $i$th observation, respectively.
We observe the triples $(Y_i, \delta_i, \X_i)$ and define
\[
Y_i = \min(T_i, C_i) \; \mbox{ and } \; \delta_i = I(T_i \leq C_i),
\]
which represent the observed response variable and the censoring indicator, respectively.
We consider the quantile regression model
\begin{equation}
T_i = \X_{i}^{T} \bb(\tau) + \epsilon_{i} (\tau),
\end{equation}
where $\beta(\tau)$ for some $\tau \in (0,1)$ is a $p$-dimensional quantile coefficient vector and $\epsilon_{i} (\tau)$ is a random error whose $\tau$th conditional quantile equals zero. The conditional quantile function is defined as
\begin{equation}
Q_{T_{i}} (\tau | \X_i) = \X_i^T \bb (\tau),
\end{equation}
where $Q_{T_{i}} (\tau | \X_i) = \inf\{t: F (t| \X_i ) \geq \tau \}$ is the $\tau$th conditional quantile of $T_i$ given $\X_i$, and $F (t| \X_i)$ is the conditional cumulative distribution function of the survival time $t$ given $\X_i$.

 Several approaches such as \cite{portnoy2003censored}, \cite{peng2008survival}, and \cite{wang2009locally} can be used to estimate the conditional quantile coefficients $\bb (\tau)$.
For instance, let us consider \cite{wang2009locally} as the basis of quantile regression for the outlier detection algorithms.
Previous methods have stringent assumptions, such as unconditional independence of the survival time and the censoring variable, or global linearity at all quantile levels.
To alleviate the assumptions, \cite{wang2009locally} proposed locally weighted censored quantile regression based on the local Kaplan-Meier estimator with Nadaraya-Watson\rq{}s type weights and a biquadratic kernel function.

The local Kaplan-Meier estimates of the distribution function $F (t|\X)$ are obtained by
\begin{equation} \label{local}
\hat{F} (t | \X) = 1 - \prod_{j=1}^{n} \Big\{1 - \frac{B_{nj} (\X)}{\sum_{k=1}^{n} I(Y_k \geq Y_j ) B_{nk}(\X)} \Big\}^{\eta_{j} (t)},
\end{equation}
where $\eta_j (t) = I(Y_j \leq t, \delta_j = 1)$ and $B_{nk}(\X)$ is a sequence of non-negative weights adding up to 1. Here, the Nadaraya-Watson type weight is employed by
\begin{equation}
B_{nk}(\x) = \frac{K(\frac{\x - \x_{k}}{h_n})}{\sum_{i=1}^{n} K(\frac{\x - \x_{k}}{h_n})},
\end{equation}
where $K$ is a density kernel function and $h_n \in \mathbb{R}^{+}$ is a bandwidth converging to zero as $n \rightarrow \infty$.
By plugging the estimator (\ref{local}) into the weight function (\ref{weight}), the estimated local weights $w_i (\hat{F})$ are obtained.
A weight function $w_{i} (F)$ is used for each censored observation as follows.
\begin{equation} \label{weight}
w_i(F) = \begin{cases}
1, & \delta_i = 1 \mbox{ or } F(C_i | \X_i ) > \tau \\
\frac{\tau - F (C_i | \X_i )}{1 - F (C_i | \X_i )}, & \delta_i = 0 \mbox{ and } F (C_i | \X_i ) < \tau ,
\end{cases}
\end{equation}
where $F (C_i | \X_i)$ is the conditional cumulative distribution function of the censoring time $C_i$ given $\X_i$.
The regression coefficient estimates $\bv (\tau)$ can be obtained by minimizing the objective function
\begin{equation}
n^{-1/2}\sum_{i=1}^{n} [w_i (\hat{F}) \rho_{\tau} (Y_i - \X_{i}^{T} \bb (\tau)) +  \{1 - w_i (\hat{F})\} \rho_{\tau} (Y^{*} - \X_{i}^{T} \bb (\tau))].
\end{equation}

%%%%%%%%%%%%%%%%%%%%%%%%%%%%%%%%%%%%%%%%%%%%%%%%%%%%%%%%%%
\subsection{Residual-based algorithm}
\qquad It is natural to consider the distance from each observation to the center to identify outliers.
Utilizing the residuals from fitting quantile regression, the \pkg{quantreg} procedure in \proglang{SAS} provides an outlier detection algorithm for uncensored data. \cite{nardi1999new} proposed the residual-based outlier detection algorithm based on Cox linear regression for censored data. It can be more effective to utilize quantile regression because it is robust to outliers. Thus, we modify the residual-based algorithm for censored data by utilizing the residuals from fitting censored quantile regression in the following manner.

Let $r_i$ be the $i$th residual defined as
\[
r_{i} = Y_{i} - Q(0.50| \X_i),
\]
where $Q(0.50| \x_i)$ is the $50$th conditional quantile for the $i$th observation by censored quantile regression. The outlier indicator for the $i$th observation, $D_{i}^{r}$, is defined as
\begin{equation}
D_{i}^{r} = \begin{cases}
		1, & \mbox{if } r_i > k_{r}\hat{\sigma} \\
		0, & \mbox{otherwise},
	\end{cases}
\end{equation}
where $k_{r}$ is a resistant parameter for controlling the tightness of cut-offs and $\hat{\sigma}$ is the corrected median of the absolute residuals. That is,
\begin{equation} \label{sas:def}
\hat{\sigma} = \mbox{median}\Big\{\frac{|r_i|}{\hat{\beta}_0}, \; i = 1, \ldots, n \Big\},
\end{equation}
where $\hat{\beta}_o = \Phi^{-1} (p)$ is the inverse cumulative distribution function (CDF) of Gaussian density with the $p$th quantile. As default values, we consider $k_r=1.5$ and $p=0.75$ like in the \proglang{SAS} procedure. The $i$th observation is declared an outlier if $D_{i}^{r} = 1$. In our \proglang{R} package \pkg{OutlierDC}, this algorithm is implemented in the function \code{odc()} with the argument \code{method = "residual"}. The algorithm is summarized as follows:\\

{\noindent {\bf Algorithm 1}: Residual-based outlier detection}
\begin{enumerate}
	\item Fit a censored quantile regression model with $\tau = 0.50$ to the data.
	\item Calculate the residuals $r_i, \; i = 1, \ldots, n$.
    \item Compute the scale parameter estimate $\hat{\sigma}$ by the residuals and the inverse CDF.
    \item Declare each observation an outlier if its corresponding residual is larger than $k_{r} \hat{\sigma}$.
\end{enumerate}

%%%%%%%%%%%%%%%%%%%%%%%%%%%%%%%%%%%%%%%%%%%%%%%%%%%%%%%%%%
\subsection{Boxplot algorithm}
\qquad A simple outlier detection approach based on a boxplot \citep{tukey1977exploratory} has widely been used for uncensored data. \cite{cho2008outlierd} and \cite{eo2012outlier} proposed to use the boxplot algorithm based on quantile regression for high-throughput high-dimensional data. We modify the boxplot algorithm using quantile regression for censored data in the following manner.

A censored quantile regression model is fitted to obtain the $25$th and $75$th conditional quantile estimates, $ Q(0.25| \X_i)$ and $ Q(0.75| \X_i)$, respectively. The inter-quantile range (IQR) for the $i$th observation can be obtained by
\[
IQR(\X_i) = Q(0.75| \X_i) - Q(0.25| \X_i).
\]
The outlier indicator for the $i$th observation, $D_{i}^{b}$, is defined as
\begin{equation}
D_{i}^{b} = \begin{cases}
		1, & \mbox{if } Y_i > UF_i \\
		0, & \mbox{otherwise},
	\end{cases}
\end{equation}
where the upper fence is defined as $UF_i =  Q(0.75| \X_i) + k_{b} IQR(\X_i)$ and $k_{b}$ is to control the tightness of cut-offs with a default value of 1.5. If an observation is located above the fence, we declare it an outlier.
The algorithm is powerful, particularly when the variability of data is heterogeneous. We implement the algorithm in the function \code{odc()} with the argument \code{method = "boxplot"}. It can be summarized as follows. \\

{\noindent {\bf Algorithm 2}: Boxplot outlier detection}
\begin{enumerate}
	\item Fit censored quantile regression models with $\tau = 0.25$ and $0.75$ to the data.
    \item Obtain the $25$th and $75$th conditional quantile estimates.
    \item Calculate $IQR(\X_i) = Q(0.75| \X_i) - Q(0.25| \X_i), i = 1,2,\ldots, n.$
    \item Construct the fence,  $UF_i =  Q(0.75| \X_i) + k_{b} IQR(\X_i), i = 1,2,\ldots, n.$
    \item Declare each observation an outlier if it is located above the fence.
\end{enumerate}

%%%%%%%%%%%%%%%%%%%%%%%%%%%%%%%%%%%%%%%%%%%%%%%%%%%%%%%%%%
\subsection{Scoring algorithm}
\qquad The residual-based and boxplot algorithms described in the previous sections focus on determining whether observations are outliers according to a threshold, which should be specified in advance.
These dichotomous algorithms, which depend solely on a pre-specified cut-off, may often be unsatisfactory.
 Moreover, the boxplot algorithm can be applicable when a single covariate exists.
Thus, we developed the scoring algorithm, which provides the outlying degree that indicates the magnitude of deviation from the distribution of observations given the covariates.
Visualizing the scores enables the flexible determination of a threshold for outlier detection. The resulting scores are free from the levels of the covariates even though the variability of the data is heterogeneous.

The outlying score is based on the relative measure of conditional quantiles. The outlying score for the $i$th observation is defined as
\begin{equation}
s_{i} = \begin{cases}
	\frac{Y_i - Q(0.50| \X_i)}{Q(0.75| \X_i) - Q(0.50| \X_i)}, & Y_i >  Q(0.50| \X_{i})\\
	\frac{Y_i - Q(0.50| \X_i)}{Q(0.25| \X_i) - Q(0.50| \X_i)}, & Y_i \leq Q(0.50| \X_{i}).
	\end{cases}
\end{equation}
The score is the difference between the distance of the observation from the $50$th quantile relative and that of the $75$th quantile, conditional on  its corresponding covariates.
Larger scores indicate higher outlying possibilities.
The normal QQ plot of the scores enables identification of outlying observations.
When the scores are visualised, a threshold $k_s$ can be determined, and an observation $i$ is declared an outlier if $s_{i} > k_s$.
The algorithm is implemented in the \code{odc} function with the argument \code{method="score"} and summarised as follows.\\

{\noindent {\bf Algorithm 3}: Scoring outlier detection}
\begin{enumerate}
	\item Fit censored quantile regression models with $\tau = 0.25, 0.50,$ and $0.75$ to the data.
  \item Obtain the $25$th, $50$th, and $75$th conditional quantile estimates.
  \item Calculate the outlying score $s_i,\; i = 1, 2, \ldots, n$.
	\item Generate the normal QQ plot of the outlying scores.
	\item Determine a threshold $k_s$ to identify outlying observations that are outside the distribution of the majority of observations.
  \item Declare each observation an outlier if its corresponding score is greater than the threshold.
\end{enumerate}

%%%%%%%%%%%%%%%%%%%%%%%%%%%%%%%%%%%%%%%%%%%%%%%%%%%%%%%%%%
\section{Implementation} \label{package}

\qquad We develop an \proglang{R} package \pkg{OutlierDC}, which is designed to detect outliers in censored data under the \proglang{R} environment system \citep{Rlang}.
The \pkg{OutlierDC} package utilizes existing \proglang{R} packages, including \pkg{methods} \citep{Rlang}, \pkg{Formula} \citep{Formula},   \pkg{survival} \citep{survival}, and \pkg{quantreg} \citep{quantreg}.
The package \pkg{methods} is adopted to provide formal structures for object-oriented programming.
\pkg{Formula} is used to manipulate the design matrix on the \code{formula} object.
The package \pkg{survival} enables the handling of survival objects by the \code{Surv} function.
The package \pkg{quantreg} provides typical censored quantile regressions.

The function \code{odc()} plays a pivotal role in outlier detection.
The usage and input arguments are as follows:
\begin{Schunk}
	\begin{Sinput}
odc <- function(formula, data,
	method = c("score", "boxplot","residual"),
	rq.model = c("Wang", "PengHuang", "Portnoy"),
	k_r = 1.5, k_b = 1.5, h = .05)
	\end{Sinput}
\end{Schunk}

\begin{itemize}
	\item \code{formula} [Formula]: a type of \code{Formula} object with a \code{survival} object on the left-hand side of the $\sim$ operator and covariate terms on the right-hand side. The survival object with its survival time and censoring status is constructed by the \code{Surv} function.
	\item \code{data} [data.frame]: a data frame with variables used in the \code{formula}. It needs at least three variables, including survival time, censoring status, and covariates.
	\item \code{method} [character]: an outlier detection method to be used. The options \code{"score"}, \code{"boxplot"}, and \code{"residual"} implement the scoring, boxplot, and residual-based algorithms, respectively. The default is \code{"score"}.
	\item \code{rq.model} [character]: a type of censored quantile regression to be used for fitting. The options \code{"Wang"}, \code{"Portnoy"}, and \code{"PengHuang"} conduct Wang and Wang's, Portnoy's, and Peng and Huang's censored quantile regression approaches, respectively. The default is \code{"Wang"}.
	\item \code{k\_r} [numeric]: a value to control the tightness of cut-offs having a default value of 1.5 for the residual-based algorithm.		
	\item \code{k\_b} [numeric]: a value to control the tightness of cut-offs having a default value of 1.5 for the boxplot algorithm.		
	\item \code{h} [numeric]: bandwidth for locally weighted censored quantile regression with a default value of 0.05.
\end{itemize}

\begin{table}
	\caption{Output slots for the \code{S4} class \code{OutlierDC}}
	\label{tab:output}
	\begin{tabular}{lll}
	\hline
	Slot & Type & Description\\
	\hline
	call & language & evaluated function call \\	
	formula & Formula & formula to be used \\
	raw.data & data.frame & data to be used for model fitting \\
	%refined.data & data.frame & the data.frame object after removing outlying observations\\
	outlier.data & data.frame & data.frame object containing outlying observation\\
	coefficients & data.frame & estimated censored quantile regression coefficient matrix \\
	%fitted.mat & matrix & the censored quantile regression fitted value matrix \\
	score & vector & outlying scores for the scoring algorithm\\
	cutoff & scalar & estimated scale parameter for the residual-based algorithm \\
	%lower & vector & lower fence vector used for the boxplot and scoring algorithms\\
	%upper & upper & upper fence vector used for the boxplot and scoring algorithms\\
	outliers & vector & logical vector to determine which observations are outliers \\
	n.outliers & numeric & number of outliers detected\\
	method & character & outlier detection method to be used\\
	rq.model & character & censored quantile regression to be used\\
	k\_r & numeric & value to be used for the tightness of cut-offs for the residual algorithm\\
	k\_b & numeric & value to be used for the tightness of cut-offs for the boxplot algorithm\\
	k\_s& numeric & threshold used for the scoring algorithm with the \code{update} function\\
	\hline
	\end{tabular}
\end{table}

The primary arguments for analysis are \code{formula} and \code{data}, which are used to specify model formula and data.
The argument \code{method} enables the choice of one of the three outlier detection algorithms described in the previous section.
The performance of the three algorithms is investigated in a simulation study in the next section.
A type of censored quantile regression can be chosen by the argument \code{rq.model}.
Portnoy's, Peng and Huang's, and Wang and Wang's approaches are provided to fit censored quantile regression in our outlier detection algorithms.
They adopt the Kaplan-Meier estimator, the Nelson-Aalen estimator, and the local Kaplan-Meier estimator, respectively.
As the values to control the tightness of cut-offs, \code{k\_r} and \code{k\_b}, increase, the sensitivity (i.e., the probability of detecting outliers correctly) becomes lower and the specificity (i.e., the probability of detecting non-outliers correctly) becomes higher, as shown in the simulation study in the next section.
As the bandwidth for locally weighted censored quantile regression, \code{h}, increases, the regression function becomes smoother.

%Figure 1 shows the work flow of the function \code{odc} to make outlier decisions as outputs from the inputs survival time, censoring status, and covariates.
%For censored quantile regression, it provides three choices: approaches of Portnoy, Peng and Huang, and Wang and Wang.
%The smoothness of the locally weighted censored quantile regression of Wang and Wang's approach can be controlled by the bandwidth \code{h}.
%The main option is to choose one of three outlier detection algorithms: residual-based, boxplot, and scoring.
%The scoring algorithm is updated with a cut-off, $k_s$, determined from the normal QQ plot of outlying scores, while the others are applied with the cut-offs, $k_r$ and $k_b$, specified in advance.
All the outputs created by \code{odc()} are stored in an object with \code{S4} class \code{OutlierDC} with different types of structures.
Table \ref{tab:output} summarises the slot structures in the \code{S4} class \code{OutlierDC}.
In addition, there are four generic functions available for the \code{OutlierDC} class: \code{coef-method}, \code{plot-method}, \code{show-method}, and \code{update-method}, as listed in Table \ref{tab:output}. These functions allow end users to observe their results more closely.

\begin{table}
	\caption{Generic functions for the \code{S4} class \code{OutlierDC}}
	\label{tab:output}
	\begin{tabular}{ll}
	\hline
	Function & Description\\
	\hline
	\code{coef} & extract model coefficient matrix including the 10th, 25th, 50th, 75th, and 90th quantile estimates.\\	
	\code{plot} & draw residual plot, scatter plot, and normal QQ plot. \\
	\code{show} & show an output of fitted object. \\
	\code{update} & update lower and upper fences for a scoring algorithm.\\
	\hline
	\end{tabular}
\end{table}

%%%%%%%%%%%%%%%%%%%%%%%%%%%%%%%%%%%%%%%%%%%%%%%%%%%%%%%%%%
\section{Simulation study}
\label{simul}
\qquad To investigate the performance of the algorithms, Monte Carlo simulation studies were conducted, accounting for censoring, based on the heterogeneous variance simulation setting of \cite{eo2012outlier}.
%We here look at the study in terms of both comparing the algorithms and finding the resistant parameter $k$.
%Since the determination of $k$ is an ongoing controversial issue, several researchers have proposed new $k$ for the labeling of outliers.
%In the boxplot criterion $k \times IQR$, \cite{hoaglin1986performance} considered $k=1.5$ and $3$ is the best choice to avoid masking problems while \cite{ingelfinger1987biostatistics} suggested the %use of $k=2$.
%More recently,  \cite{sim2005outlier} demonstrates that the choice of $k=1.5$ or $k=3$ are inappropriate for normal samples.
%In this simulation study, we demonstrate the best values for $k$ on the proposed algorithms.

\subsection{Simulation Setting}
\qquad We first generated 500 observations for a covariate $X_{i}$ from a discrete uniform distribution $DU(1,20)$ and for an error term  $\epsilon_i$ from normal distributions $N(0, \sigma_{i}^2)$, assuming
\begin{equation}
\sigma_i^2 = \exp \Big( 3 - \frac{X_{i}}{8}\Big).
\end{equation}
To generate non-outliers, survival times $T_{i}$ were obtained from the following model:
\begin{equation}
\label{eq:sim:1}
\log T_{i} = \beta_0 + \beta_1 X_i + \epsilon_i, \quad i = 1, \ldots, 480,
\end{equation}
where $(\beta_0, \beta_1) = (10, -0.3)$.
Censoring times $\log C_{i}$ were generated from $U(0, 40)$, yielding an approximate average censoring rate of 15\% and $U(0,20)$ yielding an approximate average censoring rate of 30\%.
Then, observed times were obtained by $Y_i = \min (T_i, C_i)$. In addition, we generated 20 artificial outliers from the following model:
\begin{equation}
\label{eq:sim:2}
\log T_{i} = \beta_0 + \beta_1 X_i + c \sigma_{i} + \epsilon_i^{*}, \quad i = 481, 482, \ldots, 500,
\end{equation}
where $c$ is a constant for adjusting the outlying magnitude and $\epsilon_i^{*} = \max(0, \epsilon_i)$.
We assumed $Y_i = T_i$ without censoring for the 20 outliers.
As a result, we had a data set with a total of 500 observations consisting of 480 non-outliers and 20 outliers.
This procedure was repeated 500 times independently.

To consider different magnitudes of outliers, we set the coefficient $c$ to 3, 4, or 5 in equation (\ref{eq:sim:2}), that is, $3\sigma_i, 4\sigma_i$ or $5\sigma_i$, as seen in Tables \ref{sim:linear1} and \ref{sim:linear2}.
Then, we apply each outlier detection algorithm to the data with various parameters for cut-offs: $k_r = 1.0, 1.5, 2.0, 3.0$ for the residual-based algorithm, $k_b = 0.5, 1.0, 1.5, 2.0$ for the boxplot algorithm, and $k_s = 2.0, 3.0, 4.0$ for the scoring algorithm.
Tables \ref{sim:linear1} and \ref{sim:linear2} present the accuracy, sensitivity (i.e., the probability of detecting outliers correctly), specificity (i.e., the probability of detecting non-outliers correctly), the numbers of true positives, false positives, true negatives, and false negatives, and the selected observations achieved by each algorithm with various degrees of cut-offs under different magnitudes of outliers, averaged over 500 independent experiments.

\subsection{Simulation Results}
\qquad The simulation results for identifying the outliers from $3\sigma_i$ presented in Table \ref{sim:linear1} show that as $k_r$ increased, the residual-based algorithm achieved a higher accuracy, but a lower sensitivity level, because a larger number of observations were selected, resulting in relatively more false positives.
The levels of both sensitivity and specificity were higher than 95\% when $k_r = 1.5$.
The boxplot and scoring algorithms achieved high levels of accuracy, sensitivity, and specificity when $k_b = 1.0$ and $k_s = 3.0$.
These two algorithms identified about 25 observations as outliers; about five of these were false outliers.
The residual-based algorithm identified about 40 observations as outliers when $k_r = 1.5$, resulting in about 20 false outliers.
This means that the residual-based algorithm needed to select a larger number of observations to achieve a high level of sensitivity than did the other algorithms.
When $c = 4$ (i.e., $c\sigma_i = 4\sigma_i$), the levels of both sensitivity and specificity were very high when $k_r = 2.0$, $k_b = 1.5$, and $k_s = 4.0$.
The numbers of the observations selected were 28.4, 20.8, and 20.8, on average, resulting in 8.8, 0.8, and 0.9 false outliers, respectively.
When $c = 5$ (i.e., $c\sigma_i = 5\sigma_i$), the levels of both sensitivity and specificity were very high when $k_r = 2.0$, $k_b = 2.0$, and $k_s = 4.0$.
All the results improved because the detection problem became easier, as expected.
When the censoring rates were higher (30\%), the results also had similar patterns and improved somewhat because censoring reduced the possibility of censored non-outliers being falsely detected as outliers by lowering the observed times of censored observations.

In conclusion, the residual-based algorithm had a tendency to select a relatively high number of observations to obtain a high level of sensitivity in identifying outliers. The boxplot and scoring algorithms achieved a high level of sensitivity and specificity by selecting relatively smaller numbers of observations. All the algorithms were sensitive to the choices of cut-offs. For the scoring algorithm, the choice can be assisted by enhancing the visualization function.

\begin{table}
  \caption{Simulation results with an average censoring rate of 15\% for comparing three algorithms with various degrees ($k = k_r, k_b, k_s$) of cut-offs under different magnitudes ($c\sigma_i$) of outliers. Accuracy, sensitivity, specificity, and number of true positives (TP), false positives (FP), true negatives (TN), false negatives (FN), selected observations (\#Selected) are provided. }
	\label{sim:linear1}
	\begin{tabular}{ccccccccccc}
	\hline
	$c\sigma_i$ &  Algorithm  & $k$&  Accuracy & Sensitivity & Specificity & TP & FP & TN & FN & \#Selected \\
\hline
      &           & 1.0 & 0.909  & 1.000  & 0.905  & 20.0  & 45.7  & 434.3   & ~0.0      & 65.7\\
      & Residual  & 1.5 & 0.958  & 0.973  & 0.958  & 19.5  & 20.3  & 459.7   & ~0.5      & 39.7\\
      &           & 2.0 & 0.976  & 0.827  & 0.982  & 16.5   & ~8.8  & 471.2   & ~3.5      & 25.3\\
      &           & 3.0 & 0.978  & 0.524  & 0.997  & 10.5   & ~1.6  & 478.4   & ~9.5      & 12.0\\
\cline{2-11}
      &           & 0.5 & 0.952  & 1.000  & 0.950  & 20.0  & 24.1  & 455.9   & ~0.0      & 44.1\\
$3\sigma_i$ & Boxplot   & 1.0 & 0.990  & 1.000  & 0.989  & 20.0   & ~5.1  & 474.9   & ~0.0      & 25.1\\
      &           & 1.5 & 0.988  & 0.736  & 0.998  & 14.7   & ~0.8  & 479.2   & ~5.3      & 15.6\\
      &           & 2.0 & 0.971  & 0.277  & 1.000   & ~5.5   & ~0.1  & 479.9  & 14.5       & ~5.7\\
\cline{2-11}
      &           & 2.0 & 0.953  & 1.000  & 0.951  & 20.0  & 23.3  & 456.7   & ~0.0      & 43.3\\
      & Scoring   & 3.0 & 0.990  & 0.997  & 0.990  & 19.9   & ~5.0  & 475.0   & ~0.1      & 24.9\\
      &           & 4.0 & 0.985  & 0.674  & 0.998  & 13.5   & ~0.9  & 479.1   & ~6.5      & 14.4\\
\hline
      &           & 1.0 & 0.908  & 1.000  & 0.904  & 20.0  & 45.9  & 434.1  & ~0.0      & 65.9\\
      & Residual  & 1.5 & 0.960  & 1.000  & 0.958  & 20.0  & 20.2  & 459.8  & ~0.0      & 40.2\\
      &           & 2.0 & 0.981  & 0.977  & 0.982  & 19.5   & ~8.8  & 471.2  & ~0.5      & 28.4\\
      &           & 3.0 & 0.986  & 0.734  & 0.997  & 14.7   & ~1.6  & 478.4  & ~5.3      & 16.2\\
\cline{2-11}
      &           & 0.5 & 0.952  & 1.000  & 0.950  & 20.0  & 23.8  & 456.2  & ~0.0      & 43.8\\
$4\sigma_i$ & Boxplot   & 1.0 & 0.990  & 1.000  & 0.990  & 20.0   & ~4.9  & 475.1  & ~0.0      & 24.9\\
      &           & 1.5 & 0.998  & 1.000  & 0.998  & 20.0   & ~0.8  & 479.2  & ~0.0      & 20.8\\
      &           & 2.0 & 0.996  & 0.893  & 1.000  & 17.9   & ~0.1  & 479.9  & ~2.1      & 17.9\\
\cline{2-11}
      &           & 2.0 & 0.954  & 1.000  & 0.952  & 20.0  & 23.1  & 456.9  & ~0.0      & 43.1\\
      & Scoring   & 3.0 & 0.990  & 1.000  & 0.990  & 20.0   & ~4.8  & 475.2  & ~0.0      & 24.8\\
      &           & 4.0 & 0.998  & 0.996  & 0.998  & 19.9   & ~0.9  & 479.1  & ~0.1      & 20.8\\
\hline
      &           & 1.0 & 0.909  & 1.000  & 0.905  & 20.0  & 45.5  & 434.5  & ~0.0      & 65.5\\
      & Residual  & 1.5 & 0.961  & 1.000  & 0.959  & 20.0  & 19.6  & 460.4  & ~0.0      & 39.6\\
      &           & 2.0 & 0.983  & 1.000  & 0.983  & 20.0   & ~8.4  & 471.6  & ~0.0      & 28.4\\
      &           & 3.0 & 0.992  & 0.875  & 0.997  & 17.5   & ~1.5  & 478.5  & ~2.5      & 19.0\\
\cline{2-11}
      &           & 0.5 & 0.952  & 1.000  & 0.950  & 20.0  & 24.0  & 456.0  & ~0.0      & 44.0\\
$5\sigma_i$ & Boxplot   & 1.0 & 0.990  & 1.000  & 0.990  & 20.0   & ~5.0  & 475.0  & ~0.0      & 25.0\\
      &           & 1.5 & 0.998  & 1.000  & 0.998  & 20.0   & ~0.8  & 479.2  & ~0.0      & 20.8\\
      &           & 2.0 & 1.000  & 1.000  & 1.000  & 20.0   & ~0.1  & 479.9  & ~0.0      & 20.1\\
\cline{2-11}
      &           & 2.0 & 0.953  & 1.000  & 0.951  & 20.0  & 23.5  & 456.5  & ~0.0      & 43.5\\
      & Scoring   & 3.0 & 0.990  & 1.000  & 0.990  & 20.0   & ~4.9  & 475.1  & ~0.0      & 24.9\\
      &           & 4.0 & 0.998  & 1.000  & 0.998  & 20.0   & ~0.9  & 479.1  & ~0.0      & 20.9\\
\hline
	\end{tabular}
\end{table}

\begin{table}
		\caption{Simulation results with an average censoring rate of 30\% for comparing three algorithms with various degrees ($k = k_r, k_b, k_s$) of cut-offs under different magnitudes ($c\sigma_i$) of outliers. Accuracy, sensitivity, specificity, and number of true positives (TP), false positives (FP), true negatives (TN), false negatives (FN), selected observations (\#Selected) are provided. }
	\label{sim:linear2}
	\begin{tabular}{ccccccrrrrr}
	\hline
	$c\sigma_i$ &  Algorithm  & $k$&  Accuracy & Sensitivity & Specificity & TP & FP & TN & FN & \#Selected \\
	\hline
      &           & 1.0 & 0.951  & 1.000  & 0.949  & 20.0  & 24.4  & 455.6   & 0.0      & 44.4\\
      & Residual  & 1.5 & 0.981  & 0.929  & 0.983  & 18.6   & 8.2  & 471.8   & 1.4      & 26.8\\
      &           & 2.0 & 0.985  & 0.758  & 0.995  & 15.2   & 2.6  & 477.4   & 4.8      & 17.8\\
      &           & 3.0 & 0.977  & 0.434  & 1.000   & 8.7   & 0.2  & 479.8  & 11.3       & 8.8\\
\cline{2-11}
      &           & 0.5 & 0.971  & 1.000  & 0.970  & 20.0  & 14.4  & 465.6   & 0.0      & 34.4\\
$3\sigma_i$ & Boxplot   & 1.0 & 0.994  & 0.999  & 0.994  & 20.0   & 2.8  & 477.2   & 0.0      & 22.8\\
      &           & 1.5 & 0.985  & 0.639  & 0.999  & 12.8   & 0.4  & 479.6   & 7.2      & 13.2\\
      &           & 2.0 & 0.970  & 0.250  & 1.000   & 5.0   & 0.1  & 479.9  & 15.0       & 5.1\\
\cline{2-11}
      &           & 2.0 & 0.972  & 1.000  & 0.971  & 20.0  & 14.0  & 466.0   & 0.0      & 34.0\\
      & Scoring   & 3.0 & 0.994  & 0.986  & 0.994  & 19.7   & 2.8  & 477.2   & 0.3      & 22.5\\
      &           & 4.0 & 0.982  & 0.587  & 0.999  & 11.7   & 0.6  & 479.4   & 8.3      & 12.3\\
\hline
      &           & 1.0 & 0.951  & 1.000  & 0.949  & 20.0  & 24.5  & 455.5  & 0.0      & 44.5\\
      & Residual  & 1.5 & 0.984  & 1.000  & 0.983  & 20.0   & 8.2  & 471.8  & 0.0      & 28.2\\
      &           & 2.0 & 0.992  & 0.933  & 0.995  & 18.7   & 2.6  & 477.4  & 1.3      & 21.2\\
      &           & 3.0 & 0.986  & 0.654  & 1.000  & 13.1   & 0.2  & 479.8  & 6.9      & 13.3\\
\cline{2-11}
      &           & 0.5 & 0.971  & 1.000  & 0.970  & 20.0  & 14.6  & 465.4  & 0.0      & 34.5\\
$4\sigma_i$ & Boxplot   & 1.0 & 0.995  & 1.000  & 0.994  & 20.0   & 2.7  & 477.3  & 0.0      & 22.7\\
      &           & 1.5 & 0.999  & 1.000  & 0.999  & 20.0   & 0.5  & 479.5  & 0.0      & 20.5\\
      &           & 2.0 & 0.993  & 0.824  & 1.000  & 16.5   & 0.1  & 479.9  & 3.5      & 16.6\\
\cline{2-11}
      &           & 2.0 & 0.972  & 1.000  & 0.971  & 20.0  & 14.1  & 465.9  & 0.0      & 34.1\\
      & Scoring   & 3.0 & 0.994  & 1.000  & 0.994  & 20.0   & 2.8  & 477.2  & 0.0      & 22.8\\
      &           & 4.0 & 0.998  & 0.985  & 0.999  & 19.7   & 0.6  & 479.4  & 0.3      & 20.3\\
\hline
      &           & 1.0 & 0.952  & 1.000  & 0.950  & 20.0  & 23.9  & 456.1  & 0.0      & 43.9\\
      & Residual  & 1.5 & 0.984  & 1.000  & 0.983  & 20.0   & 8.0  & 472.0  & 0.0      & 28.0\\
      &           & 2.0 & 0.995  & 0.999  & 0.995  & 20.0   & 2.5  & 477.5  & 0.0      & 22.5\\
      &           & 3.0 & 0.992  & 0.815  & 1.000  & 16.3   & 0.2  & 479.8  & 3.7      & 16.5\\
\cline{2-11}
      &           & 0.5 & 0.972  & 1.000  & 0.971  & 20.0  & 14.1  & 465.9  & 0.0      & 34.1\\
$5\sigma_i$ & Boxplot   & 1.0 & 0.995  & 1.000  & 0.995  & 20.0   & 2.6  & 477.4  & 0.0      & 22.6\\
      &           & 1.5 & 0.999  & 1.000  & 0.999  & 20.0   & 0.4  & 479.6  & 0.0      & 20.4\\
      &           & 2.0 & 1.000  & 0.999  & 1.000  & 20.0   & 0.1  & 479.9  & 0.0      & 20.1\\
\cline{2-11}
      &           & 2.0 & 0.973  & 1.000  & 0.972  & 20.0  & 13.5  & 466.5  & 0.0      & 33.5\\
      & Scoring   & 3.0 & 0.995  & 1.000  & 0.995  & 20.0   & 2.5  & 477.5  & 0.0      & 22.5\\
      &           & 4.0 & 0.999  & 1.000  & 0.999  & 20.0   & 0.4  & 479.6  & 0.0      & 20.4\\
\hline
	\end{tabular}
\end{table}

%%%%%%%%%%%%%%%%%%%%%%%%%%%%%%%%%%%%%%%%%%%%%%%%%%%%%%%%%%
\section{Case Study}
\label{ex}
\qquad In this section, We illustrate the use of \pkg{OutlierDC} for the detection of outlying observations using real data from the US SEER database system \citep{hankey1999surveillance}.
The \code{ebd} data for patients with extrahepatic cholangiocarcinoma can be obtained from the SEER website (\url{http://seer.cancer.gov}).
To call the \code{OutlierDC} package and the \code{ebd} data set,
\begin{Schunk}
\begin{Sinput}
> library(OutlierDC)
> data(ebd)
> dim(ebd)
\end{Sinput}
\begin{Soutput}
[1] 402   6
\end{Soutput}
\end{Schunk}
The data consist of 402 observations with six variables. To take a glance at the data, display the first six observations as follows:
\begin{Schunk}
\begin{Sinput}
> head(ebd)
\end{Sinput}
\begin{Soutput}
           id meta exam status time     ratio
1787 55468952    0   12      1   26 0.0000000
1788  8883016    0   12      1   11 0.0000000
1789 10647194    0   12      0  134 0.0000000
1790 16033679    2   12      1    1 0.1666667
1791 19519884    0   12      0  111 0.0000000
1792 19574077    0   12      1    8 0.0000000
\end{Soutput}
\end{Schunk}
To illustrate the outlier detection algorithms, we utilized the number of metastatic lymph nodes (called \code{meta}) as a covariate.
The response variable is the survival time in months (\code{time}), and its censoring status is denoted by \code{status}, where 0 means censored.

\begin{figure*}
	\centering
  \includegraphics[width=0.75\textwidth]{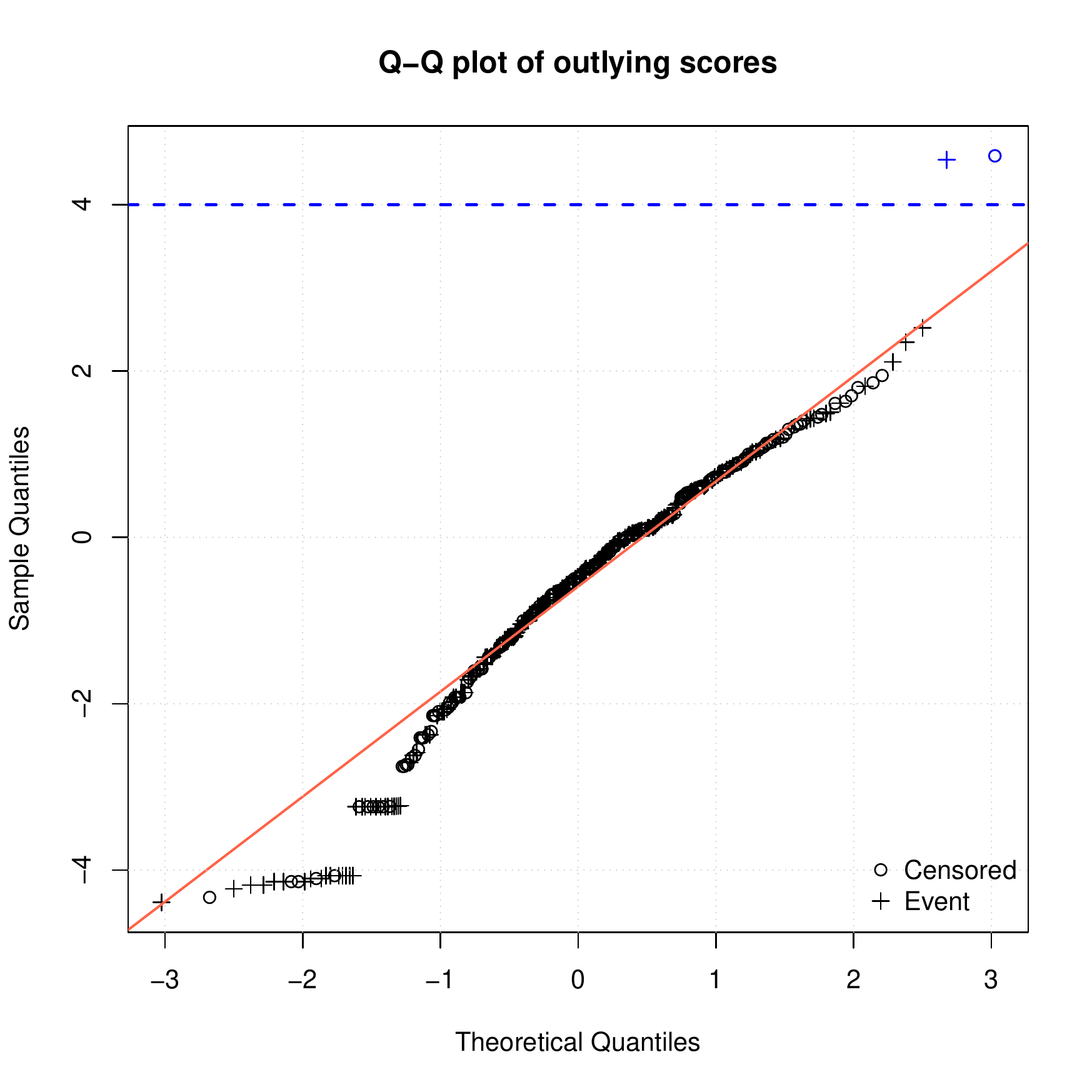}
\caption{Normal QQ plot of outlying scores with a threshold. The symbols "o" and "+" indicate the event and censored observations, respectively. The solid red line passes through the first and third  quartiles and the dotted blue line is a threshold to detect outliers. The observations above the threshold are claimed as outliers.}
\label{fig:qqplot}       % Give a unique label
\end{figure*}

The outlier detection algorithm can be run by the \code{odc} function as follows:
\begin{Schunk}
\begin{Sinput}
> fit <- odc(formula = Surv(log(time), status) ~ meta, data = ebd)
\end{Sinput}
\end{Schunk}
This command with the essential arguments \code{formula} and \code{data} runs the scoring outlier detection algorithm with Wang and Wang's censored quantile regression to create the object \code{fit}. The arguments method \code{rq.model} and \code{h} can be omitted when the defaults are used. The argument \code{k\_s} for a threshold does not need to be specified in advance for the scoring algorithm. Its full command is
\begin{Schunk}
\begin{Sinput}
fit <- odc(formula = Surv(log(time), status) ~ meta, data = ebd,
+ method = "score", h = 0.05)
\end{Sinput}
\end{Schunk}
That is, outlier detection is performed by the scoring algorithm (\code{score}) based on the locally weighted censored quantile regression with the bandwidth (\code{h = 0.05}) for selecting outliers.
The \code{show-method} command provides a summary for the \code{OutlierDC} class. To use it, type the \code{OutlierDC} object name on the \proglang{R} command line:
\begin{Schunk}
\begin{Sinput}
> fit
\end{Sinput}
\begin{Soutput}
     Outlier Detection for Censored Data

 Call: odc(formula = Surv(log(time), status) ~ meta, data = ebd)
 Algorithm: Scoring algorithm (score)
 Model: Locally weighted censored quantile regression (Wang)
 Value for cut-off k_s:
 # of outliers detected:  0

 Top 6 outlying scores:
    times delta (Intercept) meta score Outlier
346  4.48     0           1    9  4.59
327  2.71     1           1   13  4.54
326  2.08     1           1   14  2.52
296  4.86     1           1    4  2.35
354  3.09     1           1   10  2.11
233  5.29     0           1    1  1.95
\end{Soutput}
\end{Schunk}
The output via \code{show-method} consists of two parts: basic model information and top outlying scores. The first part shows the overall information such as the formula used (\code{Call}), the algorithm (\code{Algorithm}), the fitted quantile regression model (\code{Model}), the threshold value to be applied  (\code{value for cut-off k\_s}), and the number of outliers detected (\code{$\#$ of outliers detected}). The \code{Call} command displays the model formula with input arguments and the used outlier detection algorithm.
Next, the top six scores are displayed with the original data in decreasing order. The number of outliers detected (\code{\# of outliers detected}) is zero because a threshold $k_s$ (\code{Value for cut-off k\_s}) has not been provided thus far. The decision is postponed until the result is updated by the \code{update} function. A threshold can be determined by visualizing the scores. To visualize the scores,
\begin{Schunk}
\begin{Sinput}
> plot(fit)
\end{Sinput}
\end{Schunk}
The function \code{plot-method} draws a normal quantile-quantile (QQ) plot of the outlying scores, as shown in Figure \ref{fig:qqplot}. The QQ plot of outlying scores in Figure \ref{fig:qqplot} shows that the two points in the top right lie away from the line that passes through the first and third quartiles. A threshold is added by \code{k\_s} to this plot.
 Thus, the result can be updated by
\begin{Schunk}
\begin{Sinput}
> fit1 <- update(fit, k_s = 4)
> plot(fit1)
> fit1
\end{Sinput}
\begin{Soutput}
     Outlier Detection for Censored Data

 Call: odc(formula = Surv(log(time), status) ~ meta, data = ebd)
 Algorithm: Scoring algorithm (score)
 Model: Locally weighted censored quantile regression (Wang)
 Value for cut-off k_s:  4
 # of outliers detected:  2

 Top 6 outlying scores:
    times delta (Intercept) meta score Outlier
346  4.48     0           1    9  4.59       *
327  2.71     1           1   13  4.54       *
326  2.08     1           1   14  2.52
296  4.86     1           1    4  2.35
354  3.09     1           1   10  2.11
233  5.29     0           1    1  1.95
\end{Soutput}
\end{Schunk}
The two points with scores greater than the cut-off (\code{k\_s} = 4) were the 346th and 327th observations, which are marked by an asterisk.

The residual-based algorithm with a coefficient $k_r$ of 1.5 can be applied using \code{method = "residual"} with \code{k\_r = 1.5} as follows:
\begin{Schunk}
\begin{Sinput}
> fit2 <- odc(Surv(log(time), status) ~ meta, data = ebd, method = "residual", k_r = 1.5)
> plot(fit2)
> fit2
\end{Sinput}
\begin{Soutput}
     Outlier Detection for Censored Data

 Call: odc(formula = Surv(log(time), status) ~ meta, data = ebd, method = "residual",
    k_r = 1.5)
 Algorithm: Residual-based algorithm (residual)
 Model: Locally weighted censored quantile regression (Wang)
 Value for cut-off k_r:  1.5
 # of outliers detected:  9

 Outliers detected:
    times delta (Intercept) meta residual sigma Outlier
57   4.80     0           1    2     1.63   1.6       *
80   5.04     1           1    0     1.64   1.6       *
189  5.38     0           1    0     1.98   1.6       *
191  5.20     0           1    0     1.80   1.6       *
233  5.29     0           1    1     2.00   1.6       *
296  4.86     1           1    4     1.90   1.6       *
 6 of all 9 outliers were displayed.
\end{Soutput}
\end{Schunk}
Nine observations by \code{k\_r = 1.5} were selected as outliers, six of which are shown in the above output. All the outliers detected can be displayed by running \code{fit2@outlier.data}.

%\begin{figure*}
%	\centering
 % \includegraphics[width=0.75\textwidth]{residual_plot}
%\caption{plot output of the residual-based algorithm with the upper threshold. The symbols "o" and "+" indicate the event and censored observations, respectively. The dotted blue line is a threshold to detect outliers. The observations above the threshold are claimed as outliers.}
%\label{fig:residual}       % Give a unique label
%\end{figure*}
%
The boxplot algorithm with a coefficient $k_b$ of 1.5 can be applied using \code{method = "boxplot"} with \code{k\_b = 1.5}, as follows:
\begin{Schunk}
\begin{Sinput}
> fit3 <- odc(Surv(log(time), status) ~ meta, data = ebd, method = "boxplot", k_b = 1.5)
> plot(fit3)
> fit3
\end{Sinput}
\begin{Soutput}
     Outlier Detection for Censored Data

 Call: odc(formula = Surv(log(time), status) ~ meta, data = ebd, method = "boxplot",
    k_b = 1.5)
 Algorithm: Boxplot algorithm (boxplot)
 Model: Locally weighted censored quantile regression (Wang)
 Value for cut-off k_b:  1.5
 # of outliers detected:  1

 Outliers detected:
    times delta (Intercept) meta   UB Outlier
346  4.48     0           1    9 4.32       *
 1 of all 1 outliers were displayed.
\end{Soutput}
\end{Schunk}
The boxplot algorithm with a coefficient $k_b$ of 1.5 detected only one outlying point. The 346th observation detected was also detected by both the scoring and residual-based algorithms.
The boxplot algorithm with a coefficient $k_b$ of 1.0 yielded the same result as the scoring algorithm with a threshold $k_s$ of 4.0; that is, the 346th and 327th observations were detected.
%\begin{figure*}
%	\centering
  %\includegraphics[width=0.75\textwidth]{boxplot}
%\caption{plot output of the boxplot algorithm with the upper threshold. The symbols "o" and "+" indicate the event and censored observations, respectively. The censored quantile regression fits are superimposed in the black lines. The dotted blue line is a threshold to detect outliers. The observations above the threshold are claimed as outliers.}
%\label{fig:boxplot}       % Give a unique label
%\end{figure*}
%
Lastly, the \code{coef-method} function can be used to give the estimated  10th, 25th, 50th, 75th, and 90th quantile coefficients as follows:
\begin{Schunk}
\begin{Sinput}
> coef(fit)
\end{Sinput}
\begin{Soutput}
               q10    q25    q50    q75    q90
(Intercept)  1.609  2.549  3.332  4.190  5.037
meta        -0.039 -0.064 -0.091 -0.121 -0.138
\end{Soutput}
\end{Schunk}

%%%%%%%%%%%%%%%%%%%%%%%%%%%%%%%%%%%%%%%%%%%%%%%%%%%%%%%%%%
\section{Conclusion} \label{dis}
\qquad In this paper, we proposed three algorithms to detect outlying observations on the basis of censored quantile regression.
The outlier detection algorithms were implemented for censored survival data: residual-based, boxplot, and scoring algorithms.
The residual-based algorithm detects outlying observations using constant scale estimates, and therefore, it tends to select relatively many observations to achieve a high level of sensitivity in identifying outliers.
Thus, this algorithm is effective when high sensitivity is essential.
The results of our simulation study imply that the boxplot and scoring algorithms with censored quantile regression are more effective than the residual-based algorithm when considering sensitivity and specificity together.
The residual-based and boxplot algorithms require a pre-specified cut-off to determine whether observations are outliers.
Thus, these two algorithms are useful if a cut-off can be provided in advance.
Moreover, the boxplot algorithm can be applicable when a single covariate exists.
The scoring algorithm is more practical in that it provides the outlying magnitude or deviation of each point from the distribution of observations and enables the determination of a threshold by visualizing the scores; thus, this scoring algorithm is assigned as the default in our package.

All the algorithms were implemented into our developed \proglang{R} package \pkg{OutlierDC}, which is freely available via Comprehensive R Archive Network (CRAN). %It follows the Bioconductor \proglang{R} style guide to give good internal consistency.
The \code{odc} function yields the result of outlier detection by the residual-based, boxplot, or scoring algorithm.
The resulting object can be used for generic functions such as \code{show}, \code{update}, \code{plot}, and \code{coef}.
The help page for the \code{odc} function contains several examples for use in algorithms.
These can be easily accessed by the \code{example(odc)} command.
In our package, there are several options that users need to choose.
For convenience, the most effective and practical choice is assigned as the default for each option.
Thus, first-time users can run our package easily by following the illustration without a deep understanding of the presented algorithms.

%\begin{table}
%\begin{tabular}{lll}
%\hline\noalign{\smallskip}
%\end{tabular}
%\end{table}

\section{Acknowledgements}
 This research was supported by the Basic Science Research Program through the National Research Foundation of Korea (NRF) funded by the Ministry of Education, Science and Technology (2010-0007936).

%\bibliography{OutlierDC}   % name your BibTeX data base

%\begin{thebibliography}{}
% etc
%\end{thebibliography}

\end{document}